\documentstyle[12pt]{article}

\begin{document}
\hoffset = -.5truecm
\voffset = -.5truecm
\date{}
\title{\bf Breaking of Charge Independence of Nucleon-Nucleon
           Interaction
           and Bulk Properties of Nuclear Matter}
\author{{\bf G. H. Bordbar}\\
Department of Physics, Shiraz University,
Shiraz 71454, Iran\\
and\\
Institute for Studies in Theoretical Physics and Mathematics (IPM),\\
Tehran, P. O. Box 19395-5531, Iran}
\maketitle
\begin{abstract}
The effects of charge independence breaking of nucleon-nucleon interaction
on the bulk properties of nuclear matter are investigated.
Our results indicate that at high densities, the inclusion
of charge dependence in the nucleon-nucleon potential
affects the bulk properties of nuclear matter.
However, at low densities, this effect is not considerable.
It is seen that the change of our results for the nuclear matter
calculations due to the breaking of the charge independence increases by
increasing density.
It is shown that the energy contribution of the $^1S_0$
channel is sensitive to considering the charge dependence in the
nucleon-nucleon interaction.
It is indicated that the effects of charge independence breaking on the
calculated equation of state of nuclear matter are ignorable.
\end{abstract}
\newpage
\section{Introduction}
\label{intro}
Nuclear matter is defined as an uniform system of interacting nucleons
(we will consider equal number of neutrons and protons) without
electromagnetic interactions. This is a good approximation for the conditions
in the interior of a heavy nucleus. The nuclear matter is characterized
by its saturation density ($\rho = 0.17 \pm 0.02 fm^{-3}$) and energy per
nucleon ($E = -16 \pm 1 MeV$).

A central problem in nuclear many-body theory is the calculation of
properties of nuclear matter using realistic models of nucleon-nucleon
potentials. Recently, several realistic potentials are constructed which
accurately fit the proton-proton and neutron-proton scattering phase shifts.
These potentials contain terms which break charge independence \cite{1,2}.

Charge independence breaking of the nucleon-nucleon interaction means that,
in the isospin $T = 1$ state, the proton-proton ($T_z = +1$), neutron-proton
($T_z = 0$), or neutron-neutron ($T_z = -1$) interactions are different,
after electromagnetic effects have been removed. This is well established
in the $^1S_0$ state for the scattering lenghts, $a$, and effective ranges,
$r$, \cite{2},
\begin{eqnarray}
a_{pp} &=& -17.3 \pm 0.4 fm,\hskip 1truecm r_{pp} = 2.85 \pm 0.04 fm,
\nonumber\\
a_{nn} &=& -18.8 \pm 0.3 fm,\hskip 1truecm r_{nn} = 2.75 \pm 0.11 fm,
\nonumber\\
a_{np} &=& -23.75 \pm 0.01 fm,\hskip 1truecm r_{np} = 2.75 \pm 0.05 fm
\cdot
\end{eqnarray}
The charge dependence of nucleon-nucleon interaction can also be inferred
from binding energy differences of the mirror nuclei \cite{3,4,5}.

The major cause of charge independence breaking in the nucleon-nucleon
interaction is the mass difference between the charged and neutral pions
\cite{6},
\begin{eqnarray}
m_{\pi^0} &=& 134.977 MeV,\nonumber\\
m_{\pi^\pm } &=& 139.570 MeV\cdot
\end{eqnarray}
The charge dependence of nucleon-nucleon interaction due to the pion mass
difference has been calculated in references \cite{7,8}.
Whitin QCD, the charge independence breaking of nucleon-nucleon interaction
is of course due to the differences in the up and down quarks masses and
charges \cite{9,10}.

The aim of present work is the investigation of the effects of charge
independence breaking of nucleon-nucleon interaction on the properties of
nuclear matter by employing the {\it Reid-93} potential
\cite{1} in our calculations.
\section{Nuclear Matter Calculations}
\label{Sec2}
For the {\it Reid-93} potential, the proton-proton ({\it pp}),
neutron-neutron ({\it nn}), and neutron-proton ({\it np}) interactions
are different. Therefore, this potential depends on $T_z$. This shows that
in our calculations, we should consider explicitly the isospin projection
$T_z$. The procedure of the method used in our calculations is fully
discussed in references \cite{11,12,13,14,15,16,17}.

In this section, we present the results of our calculations for the various
properties of nuclear matter with inclusion of charge dependence in the
nucleon-nucleon interaction and then, we investigate the charge independence
breaking effects on our results. For this purpose, we employ the
{\it Reid-93} potential in our nuclear matter calculations in two
different cases as below.
\begin{enumerate}
\item Charge dependent case (CD-{\it Reid-93}):\\
We distinguish between the {\it pp}, {\it nn}, and {\it np}
interactions \cite{1}.
\item Charge independent case (CI-{\it Reid-93}):\\
We replace  the {\it pp} and {\it nn} interactions by the corresponding
{\it np} interaction \cite{1}. This means that we identify the
nucleon-nucleon interactions with the {\it np} interaction and therefore,
the charge dependence is removed.
\end{enumerate}
Then, we compare the results of these cases.

Our results for the total energy of nuclear matter as a function of density
with the CD-{\it Reid-93} (charge dependent case) and CI-{\it Reid-93}
(charge independent case) potentials are given in Figure 1.
It is seen that the inclusion of charge dependence in the nucleon-nucleon
interaction changes the total energy of nuclear matter, especially at high
densities. At low densities ($\rho < 0.2 fm^{-3}$), this effect is small.
In order to clarify how the charge independence breaking
(CIB) of nucleon-nucleon interaction affects the total energy of nuclear
matter, we calculate the difference between our results for the charge
dependent (CD) and charge independent (CI) cases,
\begin{equation}
\Delta E_{CIB} = E_{CD} - E_{CI}\cdot
\end{equation}
These results are presented in Table 1. We see that the change of total
energy of nuclear matter due to the breaking of the charge independence
increases by increasing density.
In Figure 1, the results of EHMMP calculations with the {\it Reid-93}
potential \cite{18} are also given for comparison. There is a good agreement
between our results and those of EHMMP \cite{18}, especially at high
densities.

In Table 2, the results of our calculations for the saturation density
$\rho_0$, energy $E(\rho_0)$, and incompressibility ${\cal K}_0$ of nuclear
matter in the charge dependent case are compared with those of charge
independent case and also with the results of EHMMP \cite{18} and
MPM \cite{19}. We see that for the saturation properies of nuclear matter,
the inclusion of charge dependence is a small effect. We also see that
our results agree with the results of others.

We know that the contribution of one-body energy to the total energy of
nuclear matter is indpendent of the nucleon-nucleon interaction.
Therefore, we investigate the effects of charge independence breaking
on the contribution of the potential energy of various channels to the
total energy. The contribution of the $^3S_1 - ^3D_1$ and $^1S_0$
channels to the potential energy are plotted versus density in Figures
2 and 3, respectively. It is seen that the $^3S_1 - ^3D_1$ energy contribution in
both cases of charge dependent and charge independent are identical.
However, a significant difference is seen for the energy contribution
of the $^1S_0$ channel in those cases. This difference increases by
increasing density.
It is also seen that the change of the energy of $^1S_0$ channel due to
the inclusion of charge dependence is repulsive.
This shows that in the channel with isospin $T = 1$, the {\it np}
interaction is more attractive than the corresponding {\it pp} and {\it nn}
interactions.
The total potential energy of nuclear matter
in the charge dependent and charge independent cases are shown in Figure 4.
We see that at high densities, the charge independence breaking of
nucleon-nucleon interaction affects the potential energy of nuclear matter.
But, at low densities, this effect is negligible.
The results of EHMMP with the {\it Reid-93} potential \cite{18}
are also shown in Figure 4. It is seen that our results are in a good
agreement with those of EHMMP \cite{18}.

The equation of state of nuclear matter is one of the most importance in
astrophysics and heavy-ion collisions. Our results for the equation of state
of nuclear matter are given in Figure 5. It can be seen that the results
of charge dependent and charge independent cases are nearly identical.
\section{Summary and Conclusion}
\label{Sec5}
The calculation of properties of nuclear matter using a realistic
nucleon-nucleon potential contained the charge dependence is of special
interest in the nuclear many-body theory. In this article, we have
calculated the various properties of nuclear matter such as the energy,
saturation properties, and equation of state using the charge dependent
{\it Reid-93} potential. At high densities, a significant difference
between our results for the energy of nuclear matter in
the case of employing charge dependent nucleon-nucleon interaction and those
of charge independent nucleon-nucleon interaction was observed.
However, at low densities, this difference is small.
This indicates that at high densities, the effects of charge independence
breaking of nucleon-nucleon interaction on the properties of nuclear matter 
are considerable.
It was seen that the energy contribution of the $^1S_0$ channel is
considerably varied by inclusion of charge dependence in the
nucleon-nucleon interaction, especially at high densities.
It was shown that the change in the equation
of state of nuclear matter due to the charge independence breaking is
negligible. Finally, the agreement between our results and those of others
was shown.
\section{Acknowledgment}
Financial support from Shiraz University research council and IPM is
gratefully acknowledged.

\newpage
\begin{table}
\caption{The change of total energy of nuclear matter due to the breaking
of charge independence of nucleon-nucleon interaction as a function of
density.
}
\label{tab1}
\begin{center}
\begin{tabular}{|c|ccccccc|}
\hline
$\rho (fm^{-3})$&0.05&0.1&0.2&0.3&0.4&0.5&0.6\\
\hline
$\Delta E_{CIB}$&0.0823&0.1416&0.2773&0.4243&0.5684&0.7061&0.8207\\
\hline
\end{tabular}
\end{center}
\end{table}
\newpage
\begin{table}
\caption{The saturation properties of nuclear matter in the cases of
charge dependent (CD) and charge independent (CI). The results of EHMMP
[18] and MPM [19] are given for comparison.}
\label{tab2}
\begin{center}
\begin{tabular}{|c|c|c|c|c|}
\hline
&$\rho_0 (fm^{-3})$&$E (\rho_0) (MeV)$&${\cal K}_0 (MeV)$\\
\hline
CD&0.272&-13.81&224\\
CI&0.277&-14.08&233\\
EHMMP&0.252&-14.45&---\\
MPM&0.271&-15.61&---\\
\hline
\end{tabular}
\end{center}
\end{table}
\newpage
\begin{figure}
\caption{The total energy of nuclear matter versus density
in the charge dependent
(full curve) and charge independent (dotted curve) cases.
The results of EHMMP [18] (Dashed curve) are given for comparison.
}
\label{fig1}
\end{figure}

\begin{figure}
\caption{As Figure 1 but for the $^3S_1 - ^3D_1$ energy contribution.
}
\label{fig2}
\end{figure}

\begin{figure}
\caption{As Figure 1 but for the $^1S_0$ energy contribution.}                                   
\label{fig3}
\end{figure}

\begin{figure}
\caption{As Figure 1 but for the total potential energy.
}
\label{fig4}
\end{figure}

\begin{figure}
\caption{As Figure 1 but for the equation of state.
}
\label{fig5}
\end{figure}
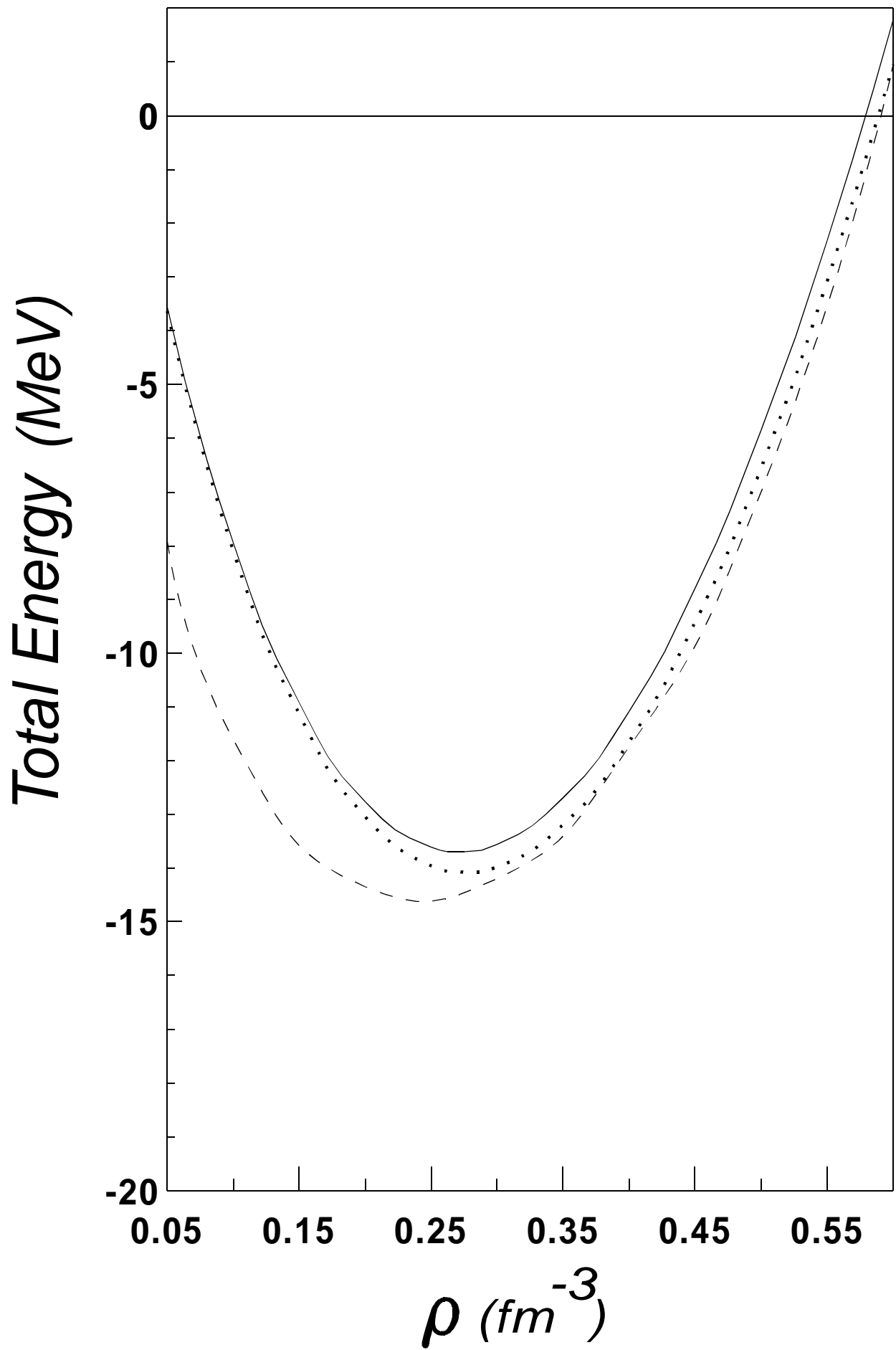
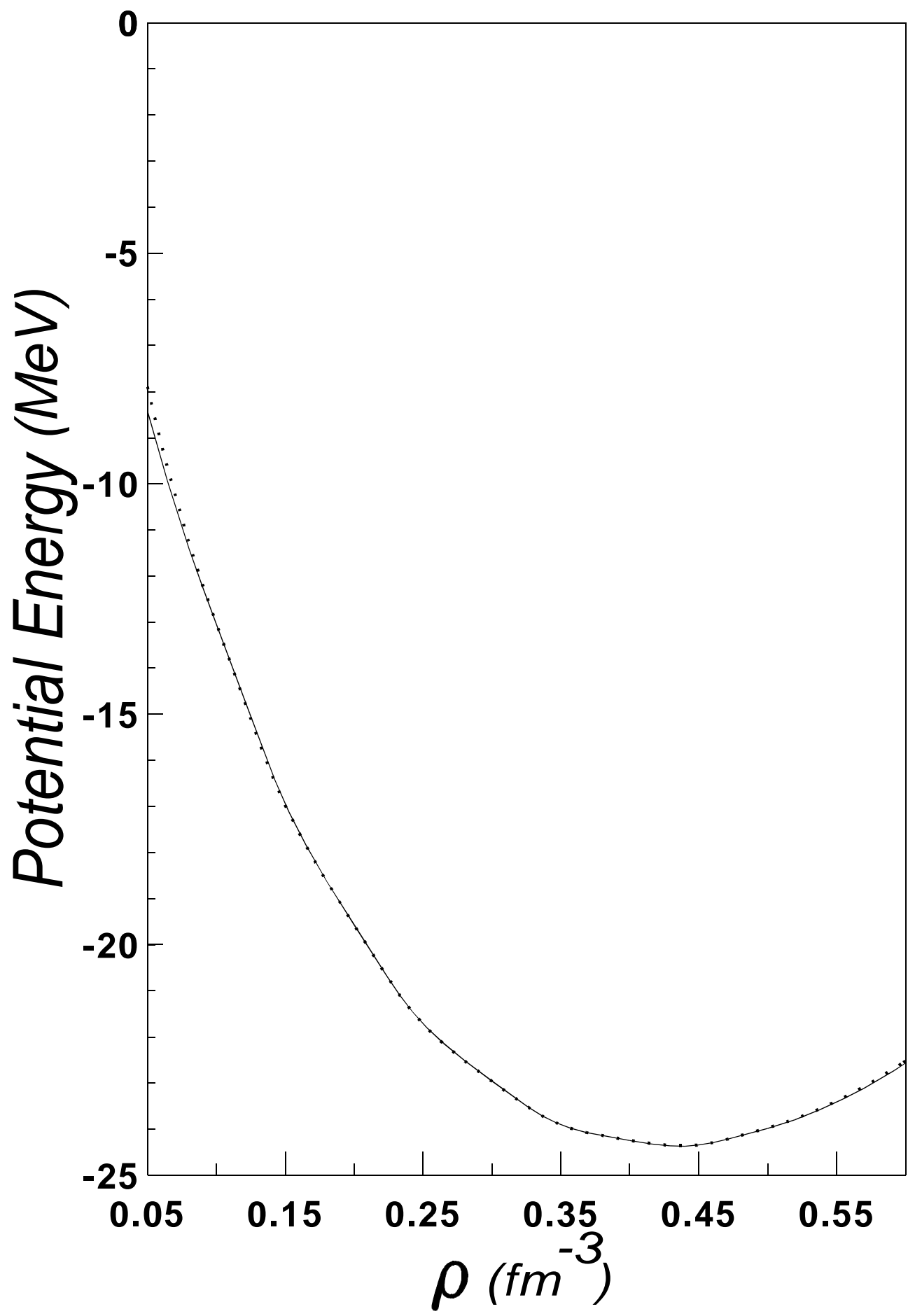
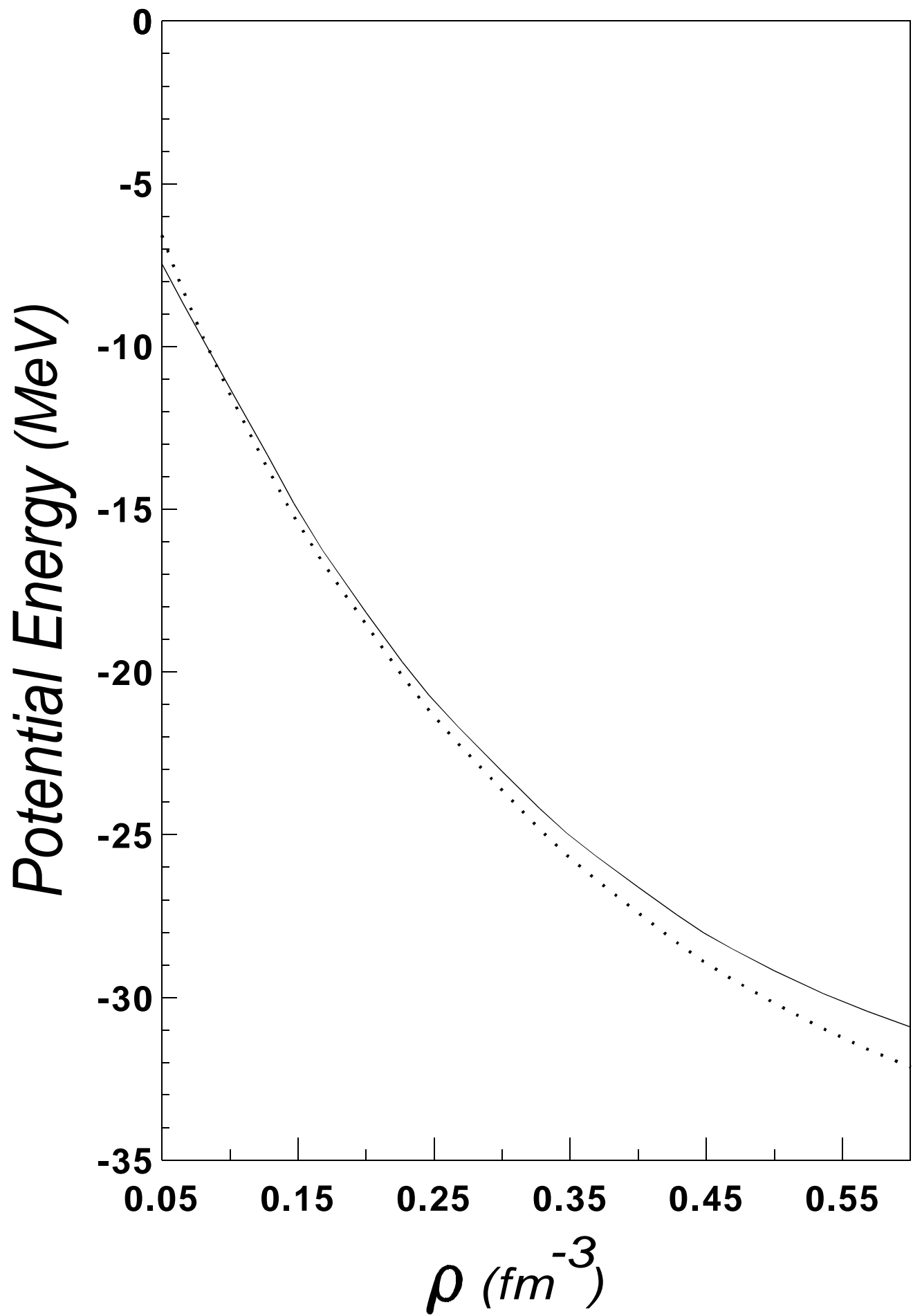
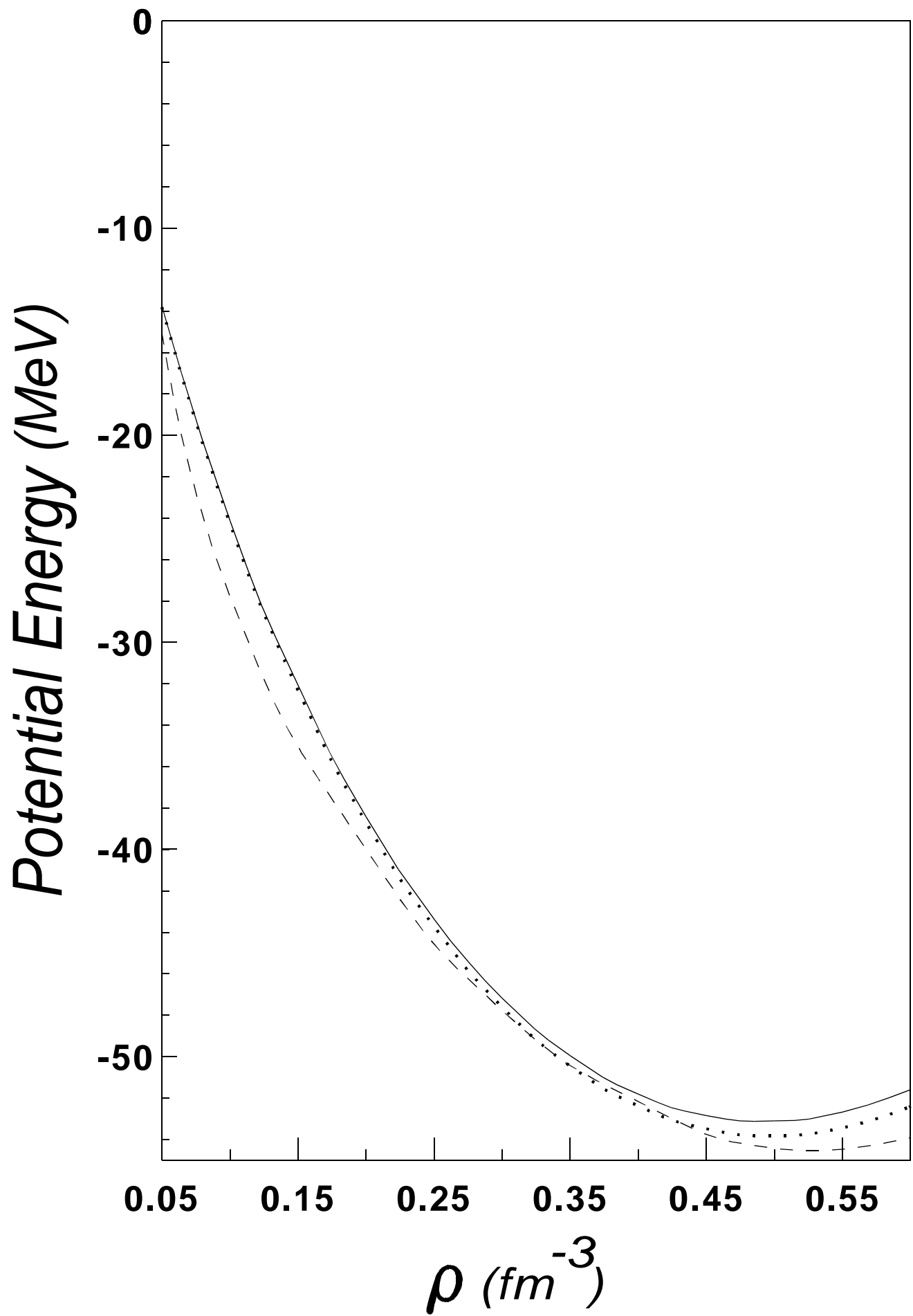
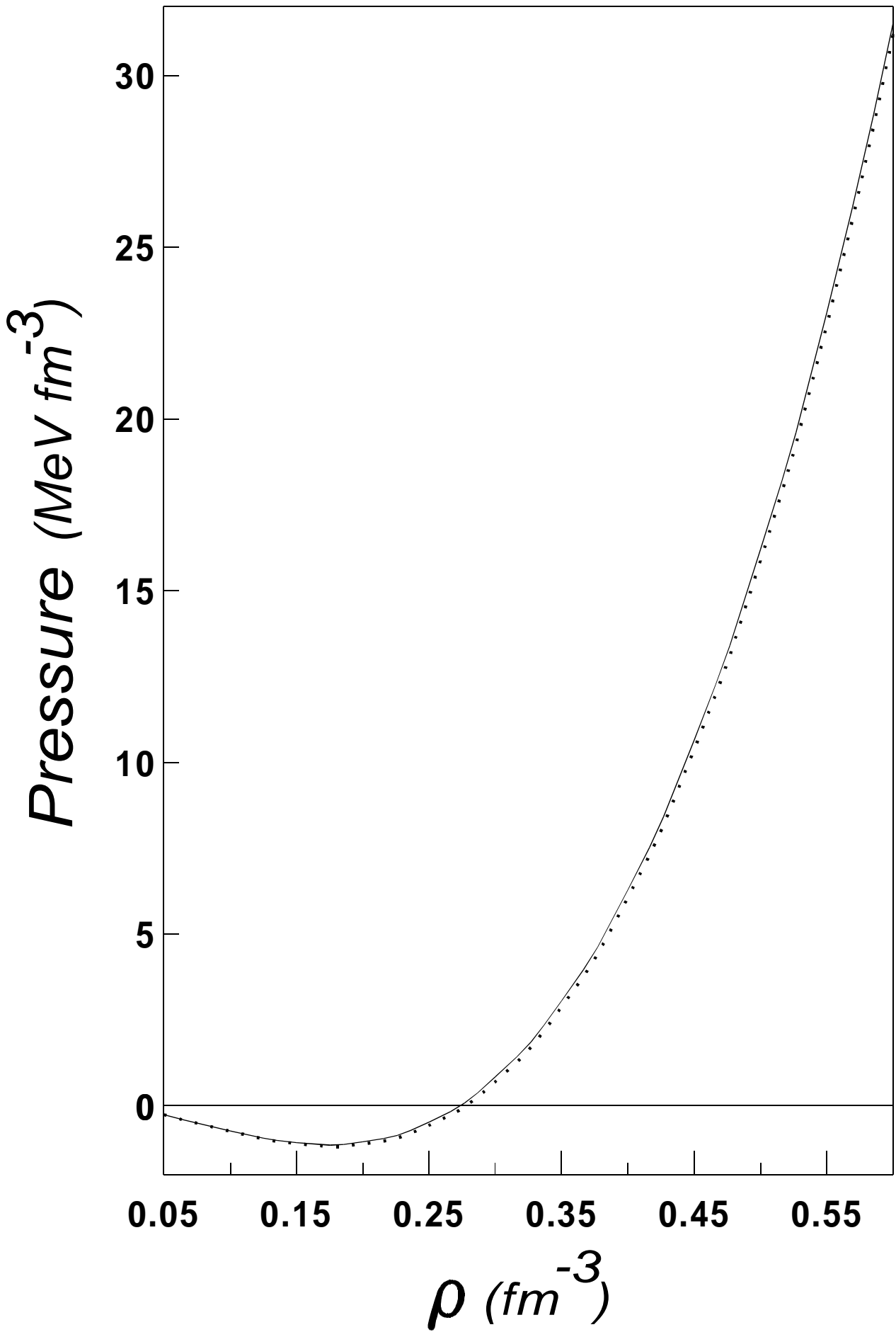
\end{document}